\documentclass[epj]{svjour}

\usepackage{graphicx}

\begin{document}

\title{%
A curious relationship between Potts glass models
}

\author{%      .
Chiaki Yamaguchi\thanks{\email{chiaki77@hotmail.com}}
}

\institute{
Kosugichou 1-359, Kawasaki 211-0063, Japan
}

\abstract{
 A Potts glass model proposed
 by Nishimori and Stephen[H.~Nishimori and M.~J.~Stephen, Phys.\ Rev.\ B \textbf{27}, 5644 (1983)]
 is analyzed by means of the replica mean field theory.
 This model is a discrete model, has a gauge symmetry,
 and is called the Potts gauge glass model.
 By comparing the present results
 with the results of the conventional Potts glass model,
 we find the coincidences and differences between the models.
 We find a coincidence that the property for the Potts glass phase in this model
 is coincident with that in the conventional model at the mean field level.
 We find a difference that,
 unlike in the case of the conventional $p$-state Potts glass model,
 this system for large $p$ does not become ferromagnetic at low temperature
 under a concentration of ferromagnetic interaction.
 The present results support the act of numerically investigating the present model
 for study of the Potts glass phase in finite dimensions.
}

\maketitle

%%%%%%%%%%%%%%%%%%
\section{Introduction} \label{sec:introduction}
%%%%%%%%%%%%%%%%%%

 The reasons for the slowing down of dynamics of supercooled liquids and the
 glass transition of the supercooled liquids to amorphous solids
 are the biggest unsolved problems in the condensed matter
 physics\cite{J1986,N1991,GS1992,N1994,N1998,BBKP2002}.
 The Potts glass model is one of abstract models for these problems\cite{BBKP2002,GGZ1980,NS1983,MMP1999,ES1983a,ES1983b,ES1983c,GKS1985,KW1987,KT1988,TK1988,CK1989,SPR1995}.

 The conventional Potts glass model\cite{ES1983a,GKS1985,SPR1995}
 does not have the gauge symmetry.
 On the other hand,
 Potts glass models, that have the gauge symmetry,
 have been proposed\cite{GGZ1980,NS1983,MMP1999}.
 The model proposed by Nishimori and Stephen \cite{NS1983} has the gauge symmetry,
 is treated in this article,
 and is called the Potts gauge glass model\cite{NS1983}.
 In Ref.\cite{NS1983},
 the two models by Nishimori and Stephen
 are proposed and have different types of exchange interactions.
 One is an extended model of the Sherrington-Kirkpatrick (SK) model\cite{SK1975,N2001}
 that an Ising model has exchange interactions of a Gaussian type.
 In this article, we call the extended model the Gaussian Potts gauge glass (GPGG) model.
 There are analytical results of the GPGG model
 by means of the replica mean field (REPMF) theory in Refs.~\cite{NS1983,C1998}.
 Another is an extended model of the bimodal Ising model\cite{N2001}
 that an Ising model has exchange interactions of a bimodal type.
 In this article, we call the extended model the present model.
 The present model is a discrete model.
 The present model has not been analyzed
 by means of the REPMF theory, and,
 in this article, the analysis by the REPMF theory is done.

 By utilizing the gauge symmetry,
 the internal energy in the present model and the upper bound of the specific heat
 in the present model are exactly calculated on a line in the phase diagram
 in any dimensions\cite{NS1983,Y2012}.
 This line is called the Nishimori line.
 The results of this model are significant 
 from the standpoint of investigation of partly exactly solvable model also.
 There are studies on this model by means of
 the numerical transfer matrix method\cite{HJPP2001,JP2002}.
 The estimation of the location of the multicritical point for this model
 based on a duality argument is also done\cite{NN2002,ONB2008}.
 A duality analysis with real-space renormalization and graph polynomials
 is recently done\cite{OJ2015}.

 In this article,
 it is shown that there is a difference between the mean field solutions
 of the GPGG model and the present model.
 It is pointed out that, when the Potts dimension $p$ goes to infinity,
 the multiple phase transition point
 of the replica symmetric (RS) approximate solution of the GPGG model
 is not on the Nishimori line\cite{NS1983,DH1988}.
 On the other hand, in this article, it is shown that
 the multiple phase transition point
 of the RS approximate solution of the present model
 is on the Nishimori line for any $p$.

 Also, by comparing the present results
 with the results of the conventional Potts glass model,
 it is found that there are the coincidences and differences between the models.
 The conventional Potts glass model is relatively-well studied
 by means of the REPMF theory\cite{ES1983a,GKS1985,SPR1995,EL1983,JK2011a,JK2011b}.
 In Ref.\cite{C1998},
 it is pointed out that the mean field solution of the GPGG model
 is coincident with the mean field solution of the conventional Potts glass model
 when the magnetization vanishes.
 In this article, it is shown that
 the mean field solution of the present model
 is coincident with the mean field solution of the conventional Potts glass model
 when the magnetization vanishes.

 For the REPMF solution of the conventional Potts glass model,
 it is known that, without an additional antiferromagnetic interaction,
 the system becomes ferromagnetic at low temperature for $p > 2$\cite{ES1983a}.
 Thus, there is a problem
 of setting the value for additional antiferromagnetic interaction
 when performing numerical analyses of the Potts glass phase in finite dimensions\cite{Cgroup2008,Bgroup2010,BKB2002a,BKB2002b,LKY2006,TH2014}, since
 the distribution of the interaction making the system no ferromagnetic
 in finite dimensions is nontrivial.
 For this problem, as done in Refs.\cite{Cgroup2008,Bgroup2010},
 there is a method of confirming the conventional Potts glass model
 with an unbiased distribution of exchange interaction for $p > 2$ in finite dimensions.
 Also, as done in Refs.\cite{BKB2002a,BKB2002b,LKY2006,TH2014},
 there are methods of using 
 antiferromagnetically biased distributions that are
 roughly estimated from the mean field solution
 in order to numerically investigate the Potts glass phase in finite dimensions.
 On the other hand, in order to numerically
 investigate the Potts glass phase in finite dimensions,
 there is a method of studying the present model
 (or the glassy Potts model mentioned below) instead of the conventional model.
 In this article, it is discussed that
 the present model does not have the problem
 of setting the value for additional antiferromagnetic interaction.

 A Potts glass model, which the magnetization does not appear, is proposed 
 by Marinari, Mossa and Parisi\cite{MMP1999}.
 This model is called the glassy Potts model.
 In the glassy Potts model, the system does not become ferromagnetic
 due to the symmetry of spins\cite{MMP1999}.
 In Ref.\cite{C1998}, it is argued that
 the Potts gauge glass model\cite{NS1983},
 which includes the present and GPGG models,
 is equivalent to the glassy Potts model.
 A detailed relation between the present model and the glassy Potts model
 is described in this article.

 The present results support the act of numerically
 investigating the present model or the glassy Potts model
 for study of the Potts glass phase in finite dimensions.

 In Ref.\cite{G1989}, for large $p$ limit, 
 it is pointed out that there is a relation
 between the conventional Potts glass model and
 a Potts glass model satisfying a gauge symmetry.
 On the other hand, in this article,
 by analyzing the mean field solutions for any $p$,
 it is pointed out that
 there is a relation between the conventional Potts glass model and the present model.
 Also, in Ref.\cite{G1989}, the magnetization is not mentioned,
 while, in this article, the magnetization in the present model is mentioned.

 The present model is explained in Section \ref{sec:model}.
 The REPMF theory is applied to this model in Section \ref{sec:application}.
 A discussion based on the gauge symmetry of this model is given
 in Section \ref{sec:gauge}.
 The comparison between the results of the infinite-range models of
 the present model and the conventional Potts glass model is made
 in Section \ref{sec:comparison}.
 The comparison between the RS approximate solutions
 of the present and GPGG models is made in Section \ref{sec:replica-symmetry}.
 A detailed relation between the present model and the glassy Potts model
 is described in Section \ref{sec:relation-PottsGaugeGlasses}.
 The concluding remarks of this article are given in Section \ref{sec:concluding-remarks}.

%%%%%%%%%%%%%%%%%%
\section{The model} \label{sec:model}
%%%%%%%%%%%%%%%%%%

 The Hamiltonian ${\cal H}$ for the present model is given by\cite{NS1983}
\begin{equation}
 {\cal H} = - J \sum_{< i j >} \eta^{({\rm P})} ( \nu_{i j} + \sigma_i - \sigma_j ) \, , \label{eq:HamilPotts1}
\end{equation}
 where $< i j >$ denotes nearest-neighbor pairs, and
 $\eta^{({\rm P})} (x)$ is defined as
\begin{equation}
\eta^{({\rm P})} (x) \equiv \sum_{r = 1}^{p - 1} e^{ \frac{2 \pi i}{p} x r} \, .
\end{equation}
 $\eta^{(P)} (x)$ takes $p - 1$ when $x \, ({\rm mod} \, p) = 0$
 and takes $- 1$ otherwise.
 $\eta^{({\rm P})} (0) = \eta^{({\rm P})} (p)  = p - 1$ for example,
 and $\sum_{x = 0}^{p - 1} \eta^{({\rm P})} (x) = 0$.
 $\sigma_i$ is the Potts spin, $\sigma_i$ is a state of the spin at site $i$,
 and $\sigma_i = 0, 1, \ldots, p - 1$.
 $p$ is the Potts dimension, and $p$ is the total number of states that one spin takes.
 $\nu_{i j}$ is a quenched variable
 related to the exchange interaction between the spins at sites $i$ and $j$,
 and $\nu_{i j} = 0, 1, \ldots, p - 1$.
 If $\nu_{i j} = 0$ for all $(i j)$ pairs,
 the model is the ferromagnetic Potts model.
 The distribution $P_{\mu} (\nu_{i j})$ for $\nu_{i j}$ is given by\cite{NS1983}
\begin{equation}
 P_{\mu} (\nu_{i j}) =
 \mu \, \delta_{\nu_{i j}, 0} + \frac{1 - \mu}{p - 1} (1 - \delta_{\nu_{i j}, 0}) \, ,
 \label{eq:Pnu}
\end{equation}
 where $\delta_{x, 0}$ is the Kronecker delta,
 and $\mu$ is a concentration of ferromagnetic interaction.
 If $\mu = 1$, the model is the ferromagnetic Potts model.
 If $\mu = \frac{1}{p}$, the distribution is given by
 $P_{\mu} (\nu_{i j}) = \frac{1}{p}$, and then
 the disorder for exchange interaction is largest.
 If $p = 2$, the model is called the Edwards-Anderson model
 and is especially called the $\pm J$ Ising spin glass model.

%%%%%%%%%%%%%%%%%%
\section{The application of the replica mean field theory} \label{sec:application}
%%%%%%%%%%%%%%%%%%

 This application is based on a method\cite{VB1985} by Viana and Bray.
 In this method, a diluted model with infinite-range interactions is treated.
 The Hamiltonian of Eq.(\ref{eq:HamilPotts1}) is rewritten as 
\begin{equation}
 {\cal H} = - \sum_{i < j} J_{i j} \, \eta^{({\rm P})} ( \nu_{i j} + \sigma_i - \sigma_j ) \, ,
 \, {\rm and}
 \, \, \, J_{i j} = 0, J \, . \label{eq:HamilPotts2}
\end{equation}
 Also, the distribution for the quenched variables $J_{i j}$ and $\nu_{i j}$
 between the spins at sites $i$ and $j$ is given by
\begin{equation}
 P (J_{i j}, \nu_{i j}) = \frac{\xi}{N} \delta_{J_{i j}, J} P_{\mu} (\nu_{i j})
 + \frac{1}{p} \biggl( 1 - \frac{\xi}{N} \biggr) \delta_{J_{i j}, 0} \, .
\end{equation}
 The distribution $P_{\mu} (\nu_{i j})$ is given in Eq.(\ref{eq:Pnu}).
 $\xi$ is the coordination number, and $N$ is the total number of spins.
 In what follows, the average for the distribution $P (J_{i j}, \nu_{i j})$
 is represented as $\langle \, \rangle_{J}$,
 and the average for the distribution $P_{\mu} (\nu_{i j})$
 is represented as $\langle \, \rangle_{\nu}$.

 We use the replica method.
 In the replica method, a relation
 $\langle \log Z \rangle_J = \lim_{n \to 0} \frac{1}{n} (\langle Z^n \rangle_J - 1)$ is used,
 where $Z$ is the partition function given by
 $Z = {\rm Tr}_{\{ \sigma_i \}} \exp ( - \beta {\cal H})$.
 $\beta$ is the inverse temperature,
 $\beta = \frac{1}{k_B T}$, $k_B$ is the Boltzmann constant,
 and $T$ is the temperature.
 We define a function $\delta^{({\rm P})} (x)$ as
\begin{equation}
 \delta^{({\rm P})} (x) \equiv \frac{1}{p} [1 + \eta^{({\rm P})} (x) ] \, .
\end{equation}
 The function $\delta^{({\rm P})} (x)$ is similar to the Kronecker delta,
 but this function is slightly different
 since $\delta^{({\rm P})} (0) = \delta^{({\rm P})} (p) = 1$ for example.
 By using the function $\delta^{({\rm P})}$, $\langle Z^n \rangle_J$ is rewritten as 
\begin{eqnarray}
 & & \langle Z^n \rangle_J 
 = {\rm Tr}_{\{ \sigma^\alpha_i \}}
 \langle \exp \{ \beta \sum_{i < j} J_{i j}
 \sum_{\alpha} [p \, \delta^{({\rm P})} (y^{\alpha}_{i j} ) - 1] \}  \rangle_{J} \nonumber \\
 &=&  {\rm Tr} \, \exp \biggl\{ \sum_{i < j} \log \biggl[ 1 + 
 \frac{\xi}{N} ( \langle e^{ \beta J \sum_{\alpha} 
 [ p \, \delta^{({\rm P})} (y^{\alpha}_{i j}  ) - 1 ] }
 \rangle_{\nu} - 1 ) \biggr] \biggr\} \nonumber \\
 &\approx& {\rm Tr} \, \exp \biggl[ \frac{\xi}{N} \sum_{i < j} 
 ( \langle e^{ \beta J \sum_{\alpha} [ p 
 \, \delta^{({\rm P})} (y^{\alpha}_{i j}  ) - 1 ] } \rangle_{\nu} - 1 ) \biggr] \nonumber \\
 &=& e^{- \frac{\xi (N - 1)}{2}} 
{\rm Tr} \, \exp \biggl\{ \frac{\xi}{N} e^{- n \beta J} \sum_{i < j} 
 \langle e^{ \beta J p \sum_{\alpha} 
 \delta^{( {\rm P})} (y^{\alpha}_{i j}  )  } \rangle_{\nu} \biggr\} \, ,
\end{eqnarray}
 where $y^{\alpha}_{i j} \equiv \nu_{i j} + \sigma^{\alpha}_i - \sigma^{\alpha}_j$,
 and $\alpha$ is the replica index.
 $\langle e^{\beta J p \sum_{\alpha} \delta^{(P)} (y_{i j}^{\alpha} ) } \rangle_{\nu}$
 is expanded as
\begin{eqnarray}
 & & \langle e^{ \beta J p \sum_{\alpha} \delta^{(P)} (y_{i j}^{\alpha} ) } \rangle_{\nu} 
 = \langle \prod_{\alpha}^{n} [ (e^{\beta J p} - 1) \delta^{(P)} (y_{i j}^{\alpha} ) + 1]
 \rangle_{\nu} \nonumber \\
  &=& 1 + (e^{\beta J p} - 1) \sum_{\alpha = 1}^n \langle \delta^{(P)} (y_{i j}^{\alpha} )
 \rangle_{\nu}
  \nonumber \\ & & + (e^{\beta J p} - 1 )^2
  \sum_{\alpha < \beta} \langle  \delta^{(P)} (y_{i j}^{\alpha} )  \delta^{(P)} (y_{i j}^{\beta} )
 \rangle_{\nu} \nonumber \\
 & & + (e^{\beta J p} - 1 )^3
  \sum_{\alpha < \beta < \gamma} \langle  \delta^{(P)} (y_{i j}^{\alpha} )  \delta^{(P)} (y_{i j}^{\beta} ) \delta^{(P)} (y_{i j}^{\gamma} )  
 \rangle_{\nu} \nonumber \\
 & & + \ldots \, . \label{eq:exp-nu}
\end{eqnarray}
 $\sum_{\alpha = 1}^{n} \langle \delta (y_{i j}^{\alpha} ) \rangle_{\nu}$
 is obtained as
\begin{eqnarray}
 & & \sum_{\alpha = 1}^{n} \langle \delta (y_{i j}^{\alpha} ) \rangle_{\nu} 
 = \frac{p \mu - 1}{p - 1} \frac{1}{p}
 \biggl[ n \nonumber \\ & &
 + \frac{1}{p} \sum_{\alpha = 1}^{n} \sum_{r}
 \eta^{(P)}  ( \sigma_i^{\alpha} - r ) \eta^{(P)}  ( \sigma_j^{\alpha} - r )  \biggr] 
 + \frac{1 - \mu}{p - 1} \, n \, ,
\end{eqnarray}
  where
 $\delta^{({\rm P})} (\sigma^{\alpha}_i - \sigma^{\alpha}_j )
  = \sum_{r = 0}^{p - 1} \delta^{({\rm P})} (\sigma^{\alpha}_i - r ) \,  \delta^{({\rm P})} (\sigma^{\alpha}_j - r)$ is used.
 $\sum_{\alpha_1 < \alpha_2 < \cdots < \alpha_k } \langle \delta^{(P)} (y_{i j}^{\alpha_1} ) \delta^{(P)} (y_{i j}^{\alpha_2} ) \cdots \delta^{(P)} (y_{i j}^{\alpha_k} ) \rangle_{\nu}$ for $k \ge 3$ is obtained as
\begin{eqnarray}
 & & \sum_{\alpha_1 < \alpha_2 < \cdots < \alpha_k }
 \langle \delta^{(P)} (y_{i j}^{\alpha_1} ) \delta^{(P)} (y_{i j}^{\alpha_2} )
 \cdots \delta^{(P)} (y_{i j}^{\alpha_k} ) \rangle_{\nu}
  \nonumber \\
 &\approx& \frac{p \mu - 1}{p - 1} \!
 \frac{1}{p^k} \! \sum_{\alpha_1 < \alpha_2 < \cdots < \alpha_k } \!
 \biggl[ 1 \nonumber \\ & &
  + \frac{1}{p} \sum_{l = 1}^k \sum_{r^{\alpha_l}} 
 \eta^{(P)}  ( \sigma_i^{\alpha_l}  -  r^{\alpha_l} )
 \eta^{(P)}  ( \sigma_j^{\alpha_l}  -  r^{\alpha_l} ) 
 \nonumber \\ & & + \frac{1}{p^2}
  \sum_{l < m} \sum_{r^{\alpha_l}} \sum_{r^{\alpha_m}}  
 \eta^{(P)}  ( \sigma_i^{\alpha_l} - r^{\alpha_l} )
 \eta^{(P)}  ( \sigma_j^{\alpha_l} - r^{\alpha_l} )
   \nonumber \\ & &
  \times \eta^{(P)}  ( \sigma_i^{\alpha_m} - r^{\alpha_m} )
  \eta^{(P)}  ( \sigma_j^{\alpha_m} - r^{\alpha_m} )  \biggr]
  \nonumber \\ & & +
 \frac{1 - \mu}{p - 1} \frac{1}{p^{k-1}} 
 \sum_{\alpha_1 < \alpha_2 < \cdots < \alpha_k } \biggl[ 1 
 \nonumber \\ & & +
 \frac{1}{p} \sum_{l < m} \sum_{r}  
 \eta^{(P)}  ( \sigma_i^{\alpha_l} - \sigma_i^{\alpha_m} + r ) 
 \eta^{(P)}  ( \sigma_j^{\alpha_l} - \sigma_j^{\alpha_m} + r )  \biggr]  
\nonumber \\
 &\approx& \frac{p \mu - 1}{p - 1} \!
 \frac{1}{p^k} \left( \begin{array}{c} n \\ k \end{array} \right) \!
 \biggl[ 1
  \! + \! \frac{1}{p} \frac{k}{n} \sum_{\alpha}
  \sum_{r} \eta^{(P)} ( \sigma_i^{\alpha} \! - \! r )
 \eta^{(P)} ( \sigma_j^{\alpha} \! - \! r ) 
 \nonumber \\ & & + \frac{1}{p^2}
 \frac{ \left( \begin{array}{c} k \\ 2 \end{array} \right)
 }{ \left( \begin{array}{c} n \\ 2 \end{array} \right) }
  \sum_{\alpha < \beta } \sum_{r} \sum_{s}
 \eta^{(P)}  ( \sigma_i^{\alpha} - r )
 \eta^{(P)}  ( \sigma_j^{\alpha} - r )
   \nonumber \\ & &
  \times \eta^{(P)}  ( \sigma_i^{\beta} - s )
  \eta^{(P)}  ( \sigma_j^{\beta} - s )  \biggr]
   + \frac{1 - \mu}{p - 1} \frac{1}{p^{k-1}}
 \left( \begin{array}{c} n \\ k \end{array} \right)
 \biggl[ 1 
 \nonumber \\ & & 
 + \frac{1}{p} \! \frac{ \left( \begin{array}{c} k \\ 2 \end{array} \right)
 }{ \left( \begin{array}{c} n \\ 2 \end{array} \right) } \!
 \sum_{\alpha < \beta } \sum_{r}  
 \eta^{(P)}  ( \sigma_i^{\alpha } \! - \! \sigma_i^{\beta } \! + r ) 
 \eta^{(P)}  ( \sigma_j^{\alpha } \! - \! \sigma_j^{\beta } \! + r ) \biggr] \,  ,
  \nonumber \\ & & \label{eq:1-k}
\end{eqnarray}
 where
 $\sum_r \eta^{(P)} (x - r)  \eta^{(P)}  (y - r) = p \, \eta^{(P)}  (x - y)$
 is used, and
 $\left( \begin{array}{c} x \\ y \end{array} \right) \equiv
 \frac{x !}{y ! (x - y) !}$.
 $\sum_{\alpha < \beta } \langle \delta^{(P)} (y_{i j}^{\alpha} ) \delta^{(P)} (y_{i j}^{\beta} ) \rangle_{\nu}$ is also obtained from Eq.~(\ref{eq:1-k}) at $k = 2$.
 Thus $\langle e^{ \beta J p \sum_{\alpha} \delta^{(P)} (y_{i j}^{\alpha} ) } \rangle_{\nu}$ is obtained as
\begin{eqnarray}
 & &  \langle e^{ \beta J p \sum_{\alpha} \delta^{(P)} (y_{i j}^{\alpha} ) } \rangle_{\nu} 
 \approx 1 + n \log \biggl( \frac{e^{\beta J p} + p - 1}{p} \biggr) \nonumber \\ & & 
 + \frac{p \mu - 1}{p - 1} \frac{1}{p} \frac{e^{\beta J p} - 1}{e^{\beta J p} + p - 1} 
  \sum_{\alpha = 1}^{n} \sum_{r = 0}^{q - 1}
 \eta^{(P)} ( \sigma_i^{\alpha} - r ) \eta^{(P)}  ( \sigma_j^{\alpha} - r )  
 \nonumber \\ & & 
 + \frac{p \mu - 1}{p - 1} \frac{1}{p^2}
 \biggl( \frac{e^{\beta J p} - 1}{e^{\beta J p} + p - 1} \biggr)^2
 \nonumber \\ & &
 \times \sum_{\alpha < \beta} \sum_{r = 0}^{q - 1} \sum_{s = 0}^{q - 1}
 \eta^{(P)}  ( \sigma_i^{\alpha} - r ) \eta^{(P)}  ( \sigma_j^{\alpha} - r ) 
  \nonumber \\ & & \times
 \eta^{(P)}  ( \sigma_i^{\beta} - s ) \eta^{(P)}  ( \sigma_j^{\beta} - s ) 
   + \frac{1 - \mu}{p - 1}  \biggl( \frac{e^{\beta J p} - 1}{e^{\beta J p} + p - 1} \biggr)^2
  \nonumber \\ & & \times
  \sum_{\alpha < \beta} \sum_{r = 0}^{q - 1}
 \eta^{(P)}  ( \sigma_i^{\alpha} - \sigma_i^{\beta} + r ) 
 \eta^{(P)}  ( \sigma_j^{\alpha} - \sigma_j^{\beta} + r ) \, ,
\end{eqnarray}
 where an approximation for the number of replicas $n$,
 $\left( \begin{array}{c} n \\ k \end{array} \right) \approx \frac{(- 1)^{k + 1}}{k} \, n$,
 is used.

 Therefore $\langle Z^n \rangle_J$ is written as
\begin{eqnarray}
 & & \langle Z^n \rangle_J
 \approx 
 \biggl[ \frac{e^{\beta J (p - 1)} + (p - 1) e^{- \beta J}}{p}
 \biggr]^{\frac{\xi (N - 1) n}{2}}
 \nonumber \\ & & \times
 {\rm Tr}_{\{ \sigma^\alpha_i \} } \exp \biggl\{  
 \frac{(p \mu - 1) w \xi}{2 (p - 1) p N} \sum_{\alpha = 1}^{n} \sum_{r = 0}^{q - 1}
  [ \sum_{i = 1}^N \eta^{(P)}  ( \sigma_i^{\alpha} - r ) ]^2 
 \nonumber \\ & &
  + \frac{(p \mu - 1) w^2 \xi}{2 (p - 1) p^2 N}
 \sum_{\alpha < \beta} \sum_{r = 0}^{q - 1} \sum_{s = 0}^{q - 1}
   [ \sum_{i = 1}^N \eta^{(P)}  ( \sigma_i^{\alpha} \! - \! r )
 \eta^{(P)}  ( \sigma_i^{\beta} \! - \! s ) ]^2 
  \nonumber \\ & &
  +  \frac{(1 - \mu) w^2 \xi}{2 (p - 1)N}
  \sum_{\alpha < \beta} \sum_{r = 0}^{q - 1}  [ \sum_{i = 1}^N 
 \eta^{(P)}  ( \sigma_i^{\alpha} - \sigma_i^{\beta} + r ) ]^2 \biggr\} \, , \label{eq:Z1}
\end{eqnarray}
and
\begin{equation}
 w \equiv \frac{e^{\beta J p} - 1}{e^{\beta J p} + p - 1} \, .
\end{equation}

 As order parameters for investigation of Potts glass phase,
 two order parameters are used at least.
 One is given by $q_{r s} = \frac{1}{N} \sum_i \langle \langle \eta (\sigma^{\alpha}_i - r) \rangle_T \langle \eta (\sigma^{\beta}_i - s ) \rangle_T \rangle_J$,
 and is used in Ref.\cite{GKS1985}, where
 $\langle \, \rangle_{T}$ is the thermal average.
 Another is given by $q^{(2)}_{r} = \frac{1}{N} \sum_i \langle \langle \langle \eta (\sigma^{\alpha}_i - \sigma^{\beta}_i - r ) \rangle^{(\sigma^{\alpha}_i )}_T \rangle^{(\sigma^{\beta}_i )}_T \rangle_J$,
 and is used in Ref.\cite{NS1983}.
 The term for $\eta^{(P)}  ( \sigma_i^{\alpha} - r ) \eta^{(P)}  ( \sigma_i^{\beta} - s )$
 of Eq.(\ref{eq:Z1}) is related to the order parameter $q_{r s}$.
 The term for $\eta^{(P)}  ( \sigma_i^{\alpha} - \sigma_i^{\beta} + r )$ 
 of Eq.(\ref{eq:Z1}) is related to the order parameter $q^{(2)}_{r}$.
 The boundaries of the glass phase by
 using these two order parameters can be agreed.
 Then, there can be a relation:
\begin{eqnarray}
 & & \frac{1}{p^2} \sum_{r = 0}^{q - 1} \sum_{s = 0}^{q - 1}
   [ \sum_{i = 1}^N \eta^{(P)}  ( \sigma_i^{\alpha} - r )
 \eta^{(P)}  ( \sigma_i^{\beta} - s ) ]^2 \nonumber \\
 &\approx& \frac{1}{p} \sum_{r = 0}^{q - 1}  [ \sum_{i = 1}^N 
 \eta^{(P)}  ( \sigma_i^{\alpha} - \sigma_i^{\beta} + r ) ]^2 \label{eq:q-q2} \, .
\end{eqnarray}
 The relation (\ref{eq:q-q2}) is also supported by
 a discussion based on the gauge symmetry.
 The discussion is given in Section \ref{sec:gauge}.
 By using Eqs.(\ref{eq:Z1}) and (\ref{eq:q-q2}),
 $\langle Z^n \rangle_J$ is obtained as
\begin{eqnarray}
 & & \langle Z^n \rangle_J
 \approx \biggl[ \frac{e^{\beta J (p - 1)} + (p - 1) e^{- \beta J}}{p}
 \biggr]^{\frac{\xi (N - 1) n}{2}}  \nonumber \\ & & \times
  {\rm Tr}_{\{ \sigma^\alpha_i \} } \exp \biggl\{ \frac{a_1}{2 N p}
 \sum_{\alpha = 1}^{n} \sum_{r = 0}^{q - 1}
  [ \sum_{i = 1}^N \eta^{(P)}  ( \sigma_i^{\alpha} - r ) ]^2 
  \nonumber \\ & &
  + \frac{a_2}{2 N p^2} \sum_{\alpha < \beta} \sum_{r = 0}^{q - 1} \sum_{s = 0}^{q - 1}
   [ \sum_{i = 1}^N \eta^{(P)}  ( \sigma_i^{\alpha} - r )
 \eta^{(P)}  ( \sigma_i^{\beta} - s ) ]^2 \biggr\} \, ,
  \nonumber \\ & &
\end{eqnarray}
\begin{equation}
 a_1 \equiv \frac{(p \mu - 1) (e^{\beta J p} - 1)}{(p - 1) (e^{\beta J p} + p - 1)} 
 \, \xi \, ,
\end{equation}
and
\begin{equation}
 a_2 \equiv \biggl( \frac{e^{\beta J p} - 1}{e^{\beta J p} + p - 1} \biggr)^2 \xi \, .
\end{equation}

 We apply the Stratonovich-Hubbard transformation.
 By using the Gaussian integral $\exp \bigl( \frac{a x^2}{2 N} \bigr)
 = \sqrt{\frac{N a}{2 \pi}} \int_{- \infty}^{\infty} d \tilde{x}$ $\exp \bigl( 
 - \frac{N a \tilde{x}^2}{2} + a x \tilde{x} \bigr)$,
 $\langle Z^n \rangle_J$ is written as
\begin{equation}
 \langle Z^n \rangle_J
 \! \approx \! \int_{- \infty}^{\infty} \!
 \prod_{\alpha} \! \prod_{r} \! d M_{\alpha r} \prod_{\alpha < \beta} \! \prod_{r} \! \prod_{s} \!
 d Q_{\alpha \beta r s} \, e^{- n N  A} \, ,
\end{equation}
\begin{eqnarray}
 A \! &\equiv& \! - \frac{\xi}{2} \log
 \biggl[ \frac{e^{\beta J (p - 1)} + (p - 1) e^{- \beta J}}{p} \biggr]
 \! + \! \frac{a_1}{2 p n} \sum_{\alpha} \sum_r ( M_{\alpha r} )^2 \nonumber \\ & & 
 + \frac{a_2}{2 p^2 n} \sum_{\alpha < \beta} \sum_r \sum_s ( Q_{\alpha \beta r s} )^2
   - \frac{1}{n} \log {\rm Tr}_{\{ \sigma^{\alpha} \}} \, e^{L} \, ,
\end{eqnarray}
and
\begin{eqnarray}
 L &\equiv&
 \frac{a_2}{p^2} \sum_{\alpha < \beta} \sum_r \sum_s Q_{\alpha \beta r s}
 \, \eta^{(P)}  ( \sigma^{\alpha} - r ) \eta^{(P)}
 ( \sigma^{\beta} - s )
 \nonumber \\  & &
 + \frac{a_1}{p} \sum_{\alpha} \sum_r M_{\alpha r}
  \, \eta^{(P)}  ( \sigma^{\alpha} - r ) 
 \label{eq:L} \, ,
\end{eqnarray}
 where $M_{\alpha r}$ is a parameter for the magnetization,
 and $Q_{\alpha \beta r s}$ is a parameter for the order of
 the Potts glass phase.
 In Eq.~(\ref{eq:L}), the site dependence of
 the spin variable $\sigma^{\alpha}_i$ is eliminated.
 By using $\langle Z^n \rangle_J \approx 1 + \max (- n N A)$ 
 and the replica method, the free energy $f$ per spin is obtained as
\begin{eqnarray}
 f  &=& - \frac{\xi}{2 \beta}
 \log \biggl[ \frac{e^{\beta J (p - 1)} + (p - 1) e^{- \beta J}}{p} \biggr]
 \nonumber \\ & & + \lim_{n \to 0} \max
 \biggl[
 \frac{a_2}{2 \beta p^2 n} \sum_{\alpha < \beta}
 \sum_r \sum_s ( Q_{\alpha \beta r s} )^2
 \nonumber \\  & &
 + \frac{a_1}{2 \beta p n} \sum_{\alpha} \sum_r ( M_{\alpha r} )^2
 - \frac{1}{\beta n} \log {\rm Tr}_{\{ \sigma^{\alpha} \}} \, e^{L} \biggr] \, .
 \label{eq:f}
 \end{eqnarray}
 The disorder for exchange interaction is largest at $\mu = \frac{1}{p}$.
 If $\mu = \frac{1}{p}$, $a_1$ gives zero,
 and then the terms for $M_{\alpha r}$ of Eq.(\ref{eq:f}) vanish.
 This means that, under a concentration of ferromagnetic interaction,
 the ferromagnetic phase does not appear at low temperature for any $p$ !
 In Section \ref{sec:replica-symmetry},
 the RS approximation of the model of Eq.(\ref{eq:f})
 is performed.

 When $J \to \frac{J}{\sqrt{N}}$ and $\xi \to N$,
 the model becomes the infinite-range model.
 Then $- \frac{\beta J \xi}{2} + \frac{\xi}{2} \log \bigl( \frac{e^{\beta J p} + p - 1}{p} \bigr) \to \frac{(\beta J)^2 (p - 1)}{4}$ and $a_2 \to (\beta J)^2$.
 Therefore, when the disorder for exchange interaction is largest ($\mu = \frac{1}{p}$),
 by using Eq.(\ref{eq:f}), 
 the free energy $f^{({\rm Inf})}$ per spin in the infinite-range model is obtained as
\begin{eqnarray}
 f^{({\rm Inf})}  &=& - \frac{\beta J^2}{4} (p - 1) - \lim_{n \to 0} \max \biggl[ \frac{1}{\beta n} \log {\rm Tr} e^{L^{({\rm Inf})}}
 \nonumber \\ & &
 - \frac{\beta J^2}{2 p^2 n} \sum_{\alpha < \beta}
 \sum_r \sum_s ( Q_{\alpha \beta r s} )^2 
 \biggr] \, ,  \label{eq:f2}
\end{eqnarray}
and
\begin{equation}
 L^{({\rm Inf})} \equiv \frac{(\beta J)^2}{p^2} \sum_{\alpha < \beta} \sum_r \sum_s Q_{\alpha \beta r s}
 \, \eta^{(P)}  ( \sigma^{\alpha} - r ) \eta^{(P)}  ( \sigma^{\beta} - s ) \, .
 \end{equation}
 This free energy and the free energy in the conventional Potts glass model
 are compared in Section \ref{sec:comparison}.

 At high temperature, the system renders $Q_{\alpha \beta r s} = M_{\alpha r} = 0$.
 Then, by using Eq.(\ref{eq:f}), the free energy $f^{({\rm High})}$ per spin is
 obtained as
\begin{equation}
 f^{({\rm High})} \! = \! - \frac{\xi}{2 \beta}
 \log \biggl[ \frac{e^{\beta J (p - 1)} \! + \!
 (p \! - \! 1) e^{- \beta J}}{p} \biggr]
 \! - \! \frac{\log (p)}{\beta} \, .
\end{equation}
 This free energy $f^{({\rm High})}$ is coincident with
 the free energy \cite{G1989} at high temperature in the random energy model.
 By using a relation $s = k_B \beta^2 \frac{\partial f}{\partial \beta}$,
 the entropy $s$ per spin is obtained as
 $s = k_B \log (p) + \frac{k_B \xi}{2}
 \log \biggl[ \frac{e^{\beta J (p - 1)} + (p - 1) e^{- \beta J}}{p} \biggr]
  - \frac{k_B \xi \beta J (p - 1) (e^{\beta J p} - 1)}{2 (e^{\beta J p} + p - 1)}$.
 By expanding the entropy $s$ for small $\beta$,
 the entropy is obtained as
 $s \approx k_B \log (p) - \frac{\xi k_B (p - 1) (\beta J)^2}{4}$.
 Thus, the Kauzmann temperature $T_K$ \cite{BBKP2002,SPR1995},
 where the entropy of the high temperature phase would vanish,
 is estimated as 
\begin{equation}
 T_K = \frac{J}{k_B} \sqrt{\frac{\xi (p - 1)}{4 \log (p)}} \, .
\end{equation}

%%%%%%%%%%%%%%%%%%
\section{A discussion based on the gauge symmetry} \label{sec:gauge}
%%%%%%%%%%%%%%%%%%

 We describe a discussion based on the gauge symmetry
 of the present model. For this model,
 the gauge transformation is performed by\cite{NS1983}
\begin{equation}
 \nu_{i j} \to \nu_{i j} + \tilde{\sigma}_i - \tilde{\sigma}_j
 \, \, \, {\rm and} \, \, \, \sigma_i \to \sigma_i - \tilde{\sigma}_i
 \, , \label{eq:GaugePotts-2} 
\end{equation}
 where $\tilde{\sigma}_i$ is
 an arbitrary value for $0, 1, \ldots, p - 1$ at site $i$.
 The condition, that the system is on the Nishimori line, is to satisfy
 a relation\cite{NS1983}:
\begin{equation}
 \beta = \frac{1}{p J} \log \biggl[ \frac{\mu (p - 1)}{1 - \mu} \biggr] \, . \label{eq:NLine}
\end{equation}
 This condition is drawn as a line in the phase diagram for $\mu$ and $\beta$.

 The internal energy $u_N$ per spin on the Nishimori line
 is exactly given by $u_N = - \frac{\xi J}{2} (\mu p - 1)$
 in any dimensions\cite{NS1983}.
 Also, as in the cases of the Ising model and the GPGG model,
 the magnetization $m_r$ is  written as
 $m_r = \frac{1}{N} \sum_{i=1}^N \langle \langle \eta (\sigma_i - r) \rangle_T \rangle_J$,
 the Potts glass order parameter $q^{(2)}_r$ is written as
 $q^{(2)}_r = \frac{1}{N} \sum_{i=1}^N \langle \langle \langle \eta (\sigma^{\alpha}_i - \sigma^{\beta}_i - r ) \rangle^{(\sigma^{\alpha}_i )}_T \rangle^{(\sigma^{\beta}_i )}_T \rangle_J$,
 and exactly $m_r = q^{(2)}_r$ on the Nishimori line in any dimensions.
 We omit the description of the derivation of the relation $m_r = q^{(2)}_r$ since
 the way of deriving this relation is to use a straightforward generalization
 of the Ising case\cite{N1981}.

 By using Eqs.(\ref{eq:Z1}) and (\ref{eq:q-q2}),
 $\langle Z^n \rangle_J$ is rewritten as
\begin{eqnarray}
 \langle Z^n \rangle_J
 &\approx& \biggl[ \frac{e^{\beta J (p - 1)} + (p - 1) e^{- \beta J}}{p}
 \biggr]^{\frac{\xi (N - 1) n}{2}}  \nonumber \\ & & \times
  {\rm Tr}_{\sigma^\alpha} \exp \biggl\{ \frac{a_1}{2 N p}
 \sum_{\alpha = 1}^{n} \sum_{r = 0}^{q - 1}
  [ \sum_{i = 1}^N \eta^{(P)}  ( \sigma_i^{\alpha} - r ) ]^2 
  \nonumber \\ & &
  + \frac{a_2}{2 N p} \sum_{\alpha < \beta} \sum_{r = 0}^{q - 1}  [ \sum_{i = 1}^N 
 \eta^{(P)}  ( \sigma_i^{\alpha} - \sigma_i^{\beta} + r ) ]^2 \biggr\} \, . \label{eq:ZGauge}
\end{eqnarray}
 The term for $\eta^{(P)}  ( \sigma_i^{\alpha} - r )$ of Eq.(\ref{eq:ZGauge})
 is related to the order parameter $m_r$, and
 the term for $\eta^{(P)}  ( \sigma_i^{\alpha} - \sigma_i^{\beta} + r )$ of Eq.(\ref{eq:ZGauge})
 is related to the order parameter $q^{(2)}_r$.
 In Eq.(\ref{eq:ZGauge}),
 when the weight for $m_r$ is equal to the weight for $q^{(2)}_r$,
 it is expected that $a_1 = a_2$.
 The relation $a_1 = a_2$ agrees with
 the condition for the Nishimori line in Eq.(\ref{eq:NLine}).
 This discussion supports that Eq.(\ref{eq:q-q2}) holds.

%%%%%%%%%%%%%%%%%%
\section{A comparison with the conventional Potts glass model} \label{sec:comparison}
%%%%%%%%%%%%%%%%%%

 The Hamiltonian ${\cal H}^{({\rm CPG})}$ for the conventional Potts glass model
 is given by\cite{ES1983a}
\begin{equation}
 {\cal H}^{({\rm CPG})} = - \sum_{< i j >} J_{i j}^{({\rm CPG})} \, \eta^{({\rm P})} (\sigma_i - \sigma_j ) \, , \label{eq:HamilPottsGlass}
\end{equation}
 where $J_{i j}^{({\rm CPG})}$ is a quenched variable,
 and the distribution \\
 $P^{({\rm CPG})} (J_{i j}^{({\rm CPG})})$ for $J_{i j}^{({\rm CPG})}$
 is given by\cite{ES1983a}
\begin{equation}
 P^{({\rm CPG})} (J_{i j}^{({\rm CPG})}) \! = \! \sqrt{\frac{N}{2 \pi J_1^2}}
 \exp \biggl[ - \frac{N}{2 J_1^2}
 \biggl( J_{i j}^{({\rm CPG})} \! - \! \frac{J_0}{N} \biggr)^2 \biggr] \, .
\end{equation}
 The infinite-range model at $p = 2$ is the SK model.

 The free energy $f^{({\rm CPG})}$ per spin in the infinite-range model of
 the conventional Potts glass model is known as
 $f^{({\rm CPG})} = - \frac{\beta J_1^2}{4} (p - 1) 
 + \lim_{n \to 0} \max \bigl\{ \frac{\beta J_1^2}{2 p^2 n} \sum_{\alpha < \beta}
 \sum_r \sum_s $ $( Q_{\alpha \beta r s}^{({\rm CPG})} )^2
 + \frac{1}{2 p n} \big[ J_0 + \frac{\beta J_1^2 (p - 2)}{2} \bigr]
 \sum_{\alpha} \sum_r ( M_{\alpha r}^{({\rm CPG})} )^2
 - $ \\ $\frac{1}{\beta n} \log {\rm Tr} e^{L^{({\rm CPG})}} \bigr\}$, and 
$L^{({\rm CPG})} \equiv
 \frac{(\beta J_1 )^2}{p^2} \sum_{\alpha < \beta} \sum_r \sum_s Q_{\alpha \beta r s}^{({\rm CPG})} $ $
 \, \eta^{(P)}  ( \sigma^{\alpha} - r ) \eta^{(P)}  ( \sigma^{\beta} - s ) 
  + \frac{\beta}{p} \big[ J_0 + \frac{\beta J_1^2 (p - 2)}{2} \bigr]
 \sum_{\alpha} \sum_r M_{\alpha r}^{({\rm CPG})} $ $ 
 \, \eta^{(P)}  ( \sigma^{\alpha} - r )$ \cite{ES1983a}.
 Surprisingly, $f^{({\rm CPG})}$ without the terms for $M_{\alpha r}$
is coincident with $f^{({\rm Inf})}$ of Eq.(\ref{eq:f2}) when $J = J_1$.
 Therefore, the behavior of spins in the Potts glass phase
 in the infinite-range model of the present model is the same as
 that of the conventional model.
 This means that the properties for the Potts glass phases in both models are coincident 
 at the mean field level.

 From the form of the free energy $f^{({\rm CPG})}$,
 it is suggested that,
 for numerical investigation of the Potts glass phase in finite dimensions,
 it is necessary that one chooses a proper negative value of $J_0$
 that depends on the values of $p$ and $\beta$.
 The definitive value of $J_0$ is nontrivial
 for study of the Potts glass phase in finite dimensions.
 On the other hand, for the present model,
 it is suggested that one chooses the proper value of $\mu$
 instead of the value of $J_0$.
 For the present model,
 the free energy is obtained in Eq.(\ref{eq:f}),
 so it is suggested that,
 in order to numerically investigate the Potts glass phase in finite dimensions,
 one investigates the properties at $a_1 = 0 \, \bigl( \mu = \frac{1}{p} \bigr)$.
 Then, the value of $\mu$ does not depend on the value of $\beta$.
 Therefore, from the analytical results by means of the REPMF theory,
 it is realized that
 to set the value of $J_0$ in the conventional Potts glass model
 is more difficult than to set the value of $\mu$ in the present model.
 In this respect, to investigate the present model
 is easier than to investigate the conventional Potts glass model.
 In addition, there is a possibility that, for the present model at $\mu = \frac{1}{p}$,
 the system is in the Potts glass phase in the ground state for $p > 2$
 in finite dimensions.
 Also, in Section \ref{sec:relation-PottsGaugeGlasses}, it is described that
 the present model at $\mu = \frac{1}{p}$ is equivalent to the glassy Potts model.

 From the comparison between the present model and
 the conventional Potts glass model, the following are understood.
 In the model of Eq.(\ref{eq:f2}), the replica symmetry ansatz
 is a poor approximation for $p \ne 2$ \cite{ES1983a,EL1983}.
 In the model of Eq.(\ref{eq:f2}), for $p > 4$,
 the Potts glass transition occurs with discontinuity of order parameter
 and without latent heat\cite{GKS1985}.
 For the analyses of the model of Eq.(\ref{eq:f2}),
 see references\cite{ES1983a,GKS1985,SPR1995,EL1983,CPR2012} 
 for example.

%%%%%%%%%%%%%%%%%%
\section{A comparison with the replica symmetric approximate solution of the Gaussian model} \label{sec:replica-symmetry}
%%%%%%%%%%%%%%%%%%

 The RS approximate solution of
 the Gaussian Potts gauge glass model (the GPGG model)
 has already been obtained\cite{NS1983}.
 By using $J^{(r)}_{i j}$ and $\lambda_i$ defined as
 $J^{(r)}_{i j} \equiv J e^{ \frac{2 \pi i}{p} \nu_{i j} r}$ and
 $\lambda_i \equiv e^{ \frac{2 \pi i}{p} \sigma_i}$,
 Eq.(\ref{eq:HamilPotts1}) is rewritten as
 ${\cal H} = - \sum_{< i j >} \sum_{r = 1}^{p - 1} J^{(r)}_{i j} \lambda^r_i \lambda^{p - r}_j$.
 The distribution $P ( J^{(r)}_{i j} )$ for $J^{(r)}_{i j}$
 is given by $P ( J^{(r)}_{i j} ) = (2 \pi J^2 )^{- \frac{(p - 1)}{2}}$ $\exp [ - \frac{1}{2 J^2} \sum_{r = 1}^{p - 1} (J^{(r)}_{i j} - J_0 )  (J^{(p - r)}_{i j} - J_0 ) ]$,
 where $J^{(p - r)}_{i j} = J^{(r) *}_{i j}$.
 $J^{(r)}_{i j}$ are variables for exchange interactions,
 and are quenched variables.
 It is assumed that $J^{(r)}_{i j}$ are complex values.
 On the other hand,
 the variables $\nu_{i j}$ for exchange interactions
 in the present model are discrete values.
 In numerical estimations,
 to process discrete values is generally easier than
 to process complex values.
 In this respect, to investigate the present model
 is generally easier than to investigate the GPGG model.

 We find the RS approximate solution of the present model.
 We assume the replica symmetry:
 $M_{\alpha r} = M_r = m \, \eta^{(P)}  ( r ) \, ( 0 \le m \le 1)$ and
 $Q_{\alpha \beta r s} = Q_{r s} = q \, \eta^{(P)}  ( r - s) \, ( 0 \le q \le 1)$,
 where $m$ is the order parameter for ferromagnetic phase,
 and $q$ is the order parameter for Potts glass phase.
 By using Eq.(\ref{eq:f}),
 the free energy $f^{({\rm RS})}$ per spin is obtained as
\begin{eqnarray}
 f^{({\rm RS})} &=& - \frac{\xi}{2 \beta}
 \log \biggl[ \frac{e^{\beta J (p - 1)} + (p - 1) e^{- \beta J}}{p} \biggr]
 + \frac{a_2 (p - 1)}{4 \beta}
 \nonumber \\  & &
 + \frac{a_1 m^2 (p - 1)}{2 \beta} 
 - \frac{a_2 (q - 1)^2 (p - 1)}{4 \beta}
 \nonumber \\  & &
 - \frac{1}{\beta} \int_{- \infty}^{\infty} \prod^{p - 1}_{r = 0}
 \biggl( \frac{d z_r}{\sqrt{2 \pi}} \, e^{- \frac{z_r^2}{2}} \biggr)
 \log B \, , \label{eq:RSf}
 \end{eqnarray}
and
\begin{equation}
 B \equiv \exp [ \sqrt{a_2 p q} z_0 + a_1 m (p - 1) ] 
+ \sum_{r = 1}^{p - 1} \exp [ \sqrt{a_2 p q} z_r  - a_1 m ] \, .
\end{equation}
 From a saddle point condition $\frac{\partial}{\partial \, m} \, f^{({\rm RS})} = 0$,
 $m$ is obtained as
\begin{equation}
 m = \frac{1}{p - 1} \int_{- \infty}^{\infty} \prod^{p - 1}_{r = 0}
 \biggl( \frac{d z_r}{\sqrt{2 \pi}} \, e^{- \frac{z_r^2}{2}} \biggr)
 \frac{C}{B} \, , \label{eq:RSm}
\end{equation}
and
\begin{eqnarray}
 C &\equiv& (p - 1) \exp [ \sqrt{a_2 p q} z_0 + a_1 m (p - 1) ]
 \nonumber \\ & &
 - \sum_{r = 1}^{p - 1} \exp [ \sqrt{a_2 p q} z_r  - a_1 m ] \, .
\end{eqnarray}
 From a saddle point condition $\frac{\partial}{\partial \, q} \, f^{({\rm RS})} = 0$,
 $q$ is obtained as
\begin{equation}
 q = \frac{p}{p - 1} \int_{- \infty}^{\infty} \prod^{p - 1}_{r = 0}
 \biggl( \frac{d z_r}{\sqrt{2 \pi}} \, e^{- \frac{z_r^2}{2}} \biggr)
 \frac{D}{B^2} - \frac{1}{p - 1}  \, , \label{eq:RSq}
\end{equation}
and
\begin{eqnarray}
 D &\equiv& \exp [ 2 \sqrt{a_2 p q} z_0 + 2 a_1 m (p - 1)] 
 \nonumber \\ & &
 + \sum_{r = 1}^{p - 1} \exp [ 2 \sqrt{a_2 p q} z_r  - 2 a_1 m ] \, .
\end{eqnarray}
 When the ferromagnetic order $m$ vanishes,
 the RS solution of the infinite-range model of Eq.(\ref{eq:RSf}) is coincident with
 those solutions of the conventional Potts glass model and the GPGG model.
 In these RS infinite-range models, the Potts glass phase transition 
 appears to be first order for $p > 6$ and
 to be second order otherwise\cite{NS1983,EL1983}.

 In the ferromagnetic phase, the system renders $q \approx m^2$.
 By expanding Eq.(\ref{eq:RSm}) for small $m$,
 we obtain $m \approx a_1 m$.
 This indicates $a_1 = 1$ at the ferromagnetic phase transition point.
 Thus, the ferromagnetic phase transition temperature $T_{\rm F}$ is
 obtained as
\begin{equation}
 T_{\rm F} = \frac{J p}{k_B \log \bigl[ \frac{(p \mu - 1) \xi + (p - 1)^2}
{(p \mu - 1) \xi - p + 1} \bigr] } \, .
\end{equation}
 In the Potts glass phase, the system renders $m = 0$.
 By expanding Eq.(\ref{eq:RSm}) for small $q$,
 we obtain $q \approx a_2 q$.
 This indicates $a_2 = 1$ at the Potts glass phase transition point.
 Thus, the Potts glass phase transition temperature $T_{\rm PG}$ is
 obtained as
\begin{equation}
 T_{\rm PG} = \frac{J p}{k_B \log \bigl( \frac{p + \sqrt{\xi} - 1}
{\sqrt{\xi} - 1} \bigr) } \, .
\end{equation}
 At the multiple phase transition point for the ferromagnetic, Potts glass and paramagnetic phases,
 there can be $a_1 = a_2 = 1$.
 Thus, the multiple phase transition point $(T^*, \mu^* )$ is
 obtained as
\begin{equation}
 (T^*, \mu^* ) = \biggl(  \frac{J p}{k_B \log \bigl( \frac{p + \sqrt{\xi} - 1}
{\sqrt{\xi} - 1} \bigr) }, \frac{1}{p} + \frac{p - 1}{\sqrt{\xi} p} \biggr) \, . \label{eq:MultiP}
\end{equation}
 The condition, that the system is on the Nishimori line, is given in Eq.(\ref{eq:NLine}),
 and by using Eq.(\ref{eq:NLine})
 it is confirmed that the phase transition point $(T^*, \mu^* )$ for any $p$ and any $\xi$
 is on the Nishimori line.
 When $p = 2$, the result of Eq.(\ref{eq:MultiP})
 agrees with the result of Ref.\cite{VB1985} for the $\pm J$ Ising model.

\begin{figure}[t]
\begin{center}
\includegraphics[width=0.80\linewidth]{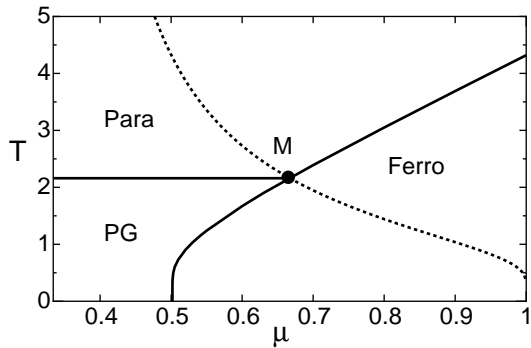}
\end{center}
\caption{
 Phase diagram of the replica symmetric approximate result
 for the diluted present model with infinite-range interactions.
 $\mu$ is the concentration of ferromagnetic interaction.
 $T$ is the temperature.
 The Potts dimension $p$ is 3, and the coordination number $\xi$ is 4. 
 The paramagnetic phase (`Para'),
 the ferromagnetic phase (`Ferro') and the Potts glass phase (`PG') are depicted.
 The Nishimori line (the dashed line) is also depicted. 
 The point `M' is the multiple phase transition point.
\label{fig:phase-diagram}
}
\end{figure}
 Fig.\ref{fig:phase-diagram} shows
 a phase diagram of the RS approximate result.
 $\mu$ is the concentration of ferromagnetic interaction.
 $T$ is the temperature.
 The Potts dimension $p$ is 3, and the coordination number $\xi$ is 4. 
 The paramagnetic phase (`Para'),
 the ferromagnetic phase (`Ferro') and the Potts glass phase (`PG') are depicted.
 The Nishimori line (the dashed line) is also depicted. 
 The point `M' is the multiple phase transition point.
 In this figure, $\frac{J}{K_B} = 1$ is set.
 The data except for the value in the ground state
 are numerically obtained from Eqs.(\ref{eq:RSm}) and (\ref{eq:RSq})
 by the recursion method.
 The value in the ground state is extrapolated from the other values.
 We can see that
 the multiple phase transition point is on the Nishimori line.

 We describe the difference between the RS approximate solutions
 of the present and GPGG models.
 We call the order parameters with no RS approximation
 $q^{({\rm True})}$ and $m^{({\rm True})}$.
 For the relationship between the Nishimori line and 
 the multiple phase transition point,
 the present RS result is different from the RS result of the GPGG model.
 It is discussed that 
 $q^{({\rm True})} = m^{({\rm True})}$ on the Nishimori line for the SK model 
 and the multiple phase transition point of
 the RS approximate solution of the SK model
 is on the Nishimori line\cite{N2001}.
 As described above,
 $q^{({\rm True})} = m^{({\rm True})}$ on the Nishimori line
 for the present model,
 and the multiple phase transition point of
 the RS approximate solution of the present model
 for any $p$ is on the Nishimori line.
 For the GPGG model,
 it is shown that $q^{({\rm True})} = m^{({\rm True})}$
 on the Nishimori line \cite{NS1983}.
 On the other hand,
 in the case of $p \to \infty$,
 the multiple phase transition point of the RS approximate solution
 of the GPGG model
 is not on the Nishimori line\cite{NS1983,DH1988}.
 Therefore,
 for the relationship between the Nishimori line and 
 the multiple phase transition point,
 the present RS result is different from the RS result of the GPGG model.
 The relations $q^{({\rm True})} = m^{({\rm True})}$ mentioned in this paragraph
 are exact relations shown by utilizing the gauge symmetries.

%%%%%%%%%%%%%%%%%%
\section{A comparison with the glassy Potts model} \label{sec:relation-PottsGaugeGlasses}
%%%%%%%%%%%%%%%%%%

 In Ref.\cite{C1998}, 
 it is argued that the Potts gauge glass model\cite{NS1983},
 which includes the present and GPGG models,
 is equivalent to the glassy Potts model\cite{MMP1999}.
 On the other hand, in Ref.\cite{C1998},
 it is not written that the ${\cal M}_{q!}$ model\cite{MMP1999} included in the glassy Potts model
 corresponds to the present model at $P_{\mu} (\nu_{i j}) = \frac{1}{p}$.
 Here, we show this.
 The Hamiltonian ${\cal H}^{({\rm GP})}$ for the glassy Potts model
 is given by\cite{MMP1999}
\begin{equation}
 {\cal H}^{({\rm GP})} = - \sum_{< i j >}  \delta_{\sigma_i, \Pi_{i j} (\sigma_j ) } \, ,
 \label{eq:HamilGP1}
\end{equation}
 where $\Pi_{i j}$ is a quenched variable between the spins at sites $i$ and $j$. 
 In the ${\cal M}_{q!}$ model,
 $\Pi_{i j}$ represents a random permutation of the Potts spin.
 Eq.(\ref{eq:HamilGP1}) is rewritten as
\begin{eqnarray}
  {\cal H}^{({\rm GP})} &=& - \sum_{< i j >}  \delta^{(p)} [ \sigma_i  - \Pi_{i j} (\sigma_j ) ] 
 \nonumber \\
 &=& -  \frac{1}{p} \sum_{< i j >} \{ \eta^{(p)} [ \sigma_i  - \Pi_{i j} (\sigma_j ) ] + 1 \} \, .
  \label{eq:HamilGP2}
\end{eqnarray}
 By using a random shift variable $\tilde{\nu}_{i j} \, (\tilde{\nu}_{i j} = 0, \ldots, p - 1)$
 instead of the random permutation $\Pi_{i j}$,
 Eq.(\ref{eq:HamilGP2}) is rewritten as
\begin{equation}
 {\cal H}^{({\rm GP})} = - \frac{1}{p} \sum_{< i j >}
 [ \eta^{(p)} ( \tilde{\nu}_{i j} + \sigma_i  - \sigma_j ) +  1 ] \, . \label{eq:HamilGP3}
\end{equation}
 By comparing Eq.(\ref{eq:HamilPotts1}) with Eq.(\ref{eq:HamilGP3}),
 it is found that 
 the ${\cal M}_{q!}$ model included in the glassy Potts model
 is essentially equivalent to the present model at $P_{\mu} (\nu_{i j}) = \frac{1}{p}$.
 Also, when $P_{\mu} (\nu_{i j}) = \frac{1}{p}$,
 there is a relation $\mu = \frac{1}{p}$.

%%%%%%%%%%%%%%%%%%
\section{Concluding remarks} \label{sec:concluding-remarks}
%%%%%%%%%%%%%%%%%%

 For the infinite-range models, it was shown that
 the Potts glass phase in the conventional Potts glass model
 is equivalent to the Potts glass phase in the present model.
 Also, it was discussed that, when performing numerical analyses,
 the conventional Potts glass model has
 the problem of setting the value for additional antiferromagnetic interaction,
 while the present model does not have the problem.
 It is expected that 
 the numerical estimations for the present model in finite dimensions
 help to understand the Potts glass phase in finite dimensions.

 For the relationship between the Nishimori line and 
 the multiple phase transition point,
 it was shown that the present replica symmetric result is different from
 the replica symmetric result of the Gaussian model (the GPGG model)
 when the Potts dimension $p$ goes to infinity.
 It seems that the difference between these results
 is caused by the difference of exchange interactions.
 This detailed investigation is a task for the future.

\end{document}